\documentstyle[11pt,aussois2003,twoside,epsf]{article}

\newcommand{\acta}[2]{{\it Acta Astron}, {\bf #1}, \rm #2}
\newcommand{\al}[2]{{\it Astron Lett}, {\bf #1}, \rm #2}
\newcommand{\ar}[2]{{\it Astron Reports}, {\bf #1}, \rm #2}
\def\Dos{D_{\rm os}}
\def\Dls{D_{\rm ls}}
\def\Dol{D_{\rm ol}}
\def\Msun{M_{\odot}}
\def\msun{$\Msun$}
\def\AU{{\rm AU}}
\def\RE{R_{\rm E}}
\def\tE{t_{\rm E}}
\def\RtE{{\tilde R}_{\rm E}}
\def\RhE{{\hat R}_{\rm E}}

\def\btheta{\mbox{\boldmath $\theta$}}
\def\thetaE{\theta_{\rm E}}
\def\thetaI{\theta_{\rm I}}
\def\thetaS{\theta_{\rm S}}
\def\pirel{\pi_{\rm rel}}

\def\kms{\;{\rm kms}^{-1}}

\def\vp{v_{\rm p}}
\def\max{\rm max}
\def\spose#1{\hbox to 0pt{#1\hss}}
\def\lta{\mathrel{\spose{\lower 3pt\hbox{$\sim$}} \raise
2.0pt\hbox{$<$}}}
\def\gta{\mathrel{\spose{\lower 3pt\hbox{$\sim$}} \raise
2.0pt\hbox{$>$}}}
\def\edcomment#1{\iffalse\marginpar{\raggedright\sl#1\/}\else\relax\fi}
\reversemarginpar
\begin{document}
%
\title{The First Heroic Decade of Microlensing}
%
\author{N.W. Evans}
\affil{Institute of Astronomy, Madingley Rd, Cambridge, CB3 0HA, UK}
\label{page:first}
\begin{abstract}
We describe the fundamentals of vanilla and exotic microlensing.
Deviations from the standard form of an achromatic, time-symmetric
lightcurve can be caused by the parallax and xallarap effects, finite
source sized effects and binarity. Three applications of microlensing
from the First Heroic Decade are reviewed in detail -- namely (i)
searches for compact dark objects in the Galactic halo, (ii) probes of
the baryonic mass distribution in the inner Galaxy and the Andromeda
Galaxy and (iii) studies of the limb darkening of source stars.
Finally, we suggest four projects for the Second Heroic Decade -- (i)
K band microlensing towards the Bulge, (ii) pixel lensing towards the
low luminosity spiral galaxy M33, (iii) polarimetry of on-going
microlensing events and (iv) astrometric microlensing with the {\it
GAIA} satellite.
\end{abstract}
\section{Introduction}

This is the end of the First Heroic Decade of Gravitational
Microlensing.  The pioneering experiments of MACHO, EROS and OGLE
reported first candidate events exactly ten years ago (Alcock et
al. 1993; Aubourg et al. 1993; Udalski et al. 1993). This led to a
frenzied outburst of activity that is only just now beginning to
subside.  After a decade of glorious achievement, two of the original
collaborations (MACHO and EROS) are winding up. Now is an opportune
moment to summarise the achievements of the First Heroic Decade, as
well as to speculate on what is needed to make the forthcoming decade
just as grand!

Historically, the discovery of microlensing at high optical depth
predates that of microlensing at low optical depth. The first-ever
event that was recognised as microlensing was the bump in the
lightcurve of image A of the Einstein Cross (Irwin et al. 1989;
Corrigan et al. 1991). The importance of microlensing at low optical
depth was generally realised only after a visionary publication of
Paczy\'nski's (1986), in which the idea of monitoring stellar images
in the Large Magellanic Cloud was first convincingly mooted. At
outset, the impetus for the microlensing surveys was the dark
matter problem. Many compact dark matter candidates in the Galactic
halo would betray their presence through microlensing. The great
achievement of MACHO and EROS has been to rule out most forms of
compact dark matter as the dominant contributors to the dark halo.
The richness of the microlensing phenomenon has led to applications
above and beyond the original aims of the surveys. These include
probes of galactic structure and stellar populations, delineation of
the Galactic bar, studies of limb darkening and planet searching. The
next decade will surely see these applications centre-stage.

\section{The Fundamentals}

\subsection{Vanilla Microlensing}

\begin{figure}
\plotone{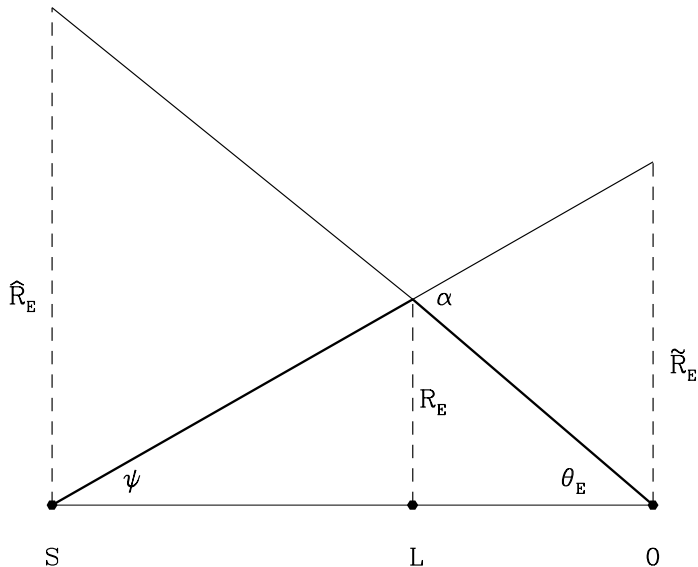}
\caption{The source (S), lens (L) and observer (O) are all aligned, so
the observer sees a bright Einstein ring of radius $\RE$. Its size
projected onto the observer's plane is $\RtE$ and onto the source
plane is $\RhE$.}
\label{fig:aligned}
\end{figure}
\begin{figure}
\plotone{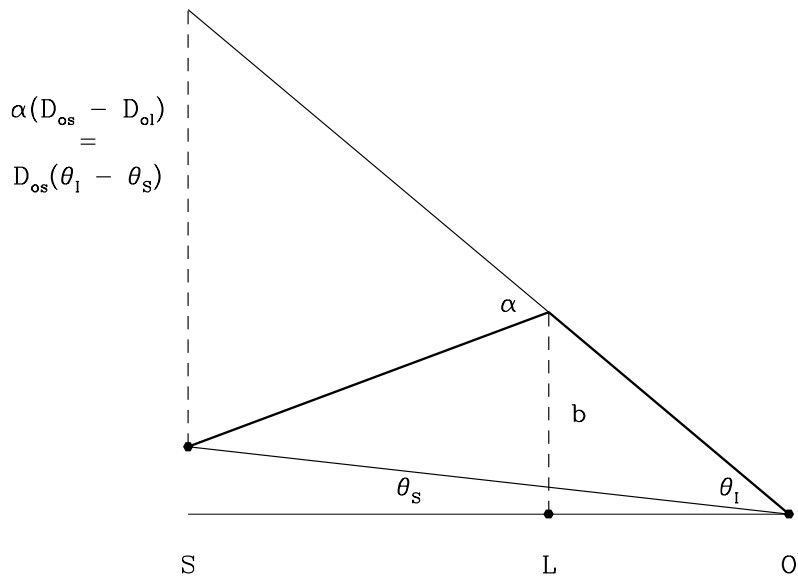}
\caption{The source (S), lens (L) and observer (O) are now misaligned,
so the observer sees two micro-images, one of which is depicted
here. (The second micro-image lies on the other side of the lens).}
\label{fig:misaligned}
\end{figure}
Suppose photons impinge upon a nearby mass $M$ with impact parameter
$b$.  Then, the General Theory of Relativity predicts that the photons
are deflected through an angle $\alpha$ given by (e.g., Landau \&
Lifshitz 1971, section 98)
\begin{equation}
\alpha = {4GM \over b c^2}.
\label{eq:einstein}
\end{equation}
Fig.~1 shows the case when the background source of radiation (S),
the lens (L) and the observer (O) are in exact alignment. Of course,
the horizontal scale of the figure is much compressed in comparison to
the vertical scale, so that the deflection $\alpha$ is minute. The
distance between observer and lens is denoted by $\Dol$, that between
observer and source by $\Dos$, and that between lens and source by
$\Dls$.  Shown in bold is the path of photons from S, which are
deflected by the lens L through an angle $\alpha$. As the figure is
axisymmetric about the optic axis joining observer and lens, the
observer sees a bright ring with radius $\RE$ given by
\begin{equation}
\RE^2 = {4GM\over c^2}{\Dol (\Dos-\Dol) \over  \Dos}.
\end{equation}
In microlensing, the angular size of the Einstein ring $\thetaE = \RE
/\Dol$ is typically of the order of microarcseconds. In some
applications, it is helpful to consider the size of the Einstein
radius projected onto the observer's plane, which is $\RtE = \RE
\Dos/\Dls$, and onto the source plane, which is $\RhE = \RE
\Dos/\Dol$.

Directly from Fig.~1, we observe that $\alpha/\RtE = \thetaE/
\RE$. So, using the Einstein deflection angle formula~(1), we deduce
that
\begin{equation}
\thetaE \RtE = \alpha \RE = {4GM\over c^2}.
\end{equation}
Also straight from Fig.~1, we observe that $\thetaE = \alpha - \psi =
\RtE/\Dol - \RtE/\Dos$, so
\begin{equation}
{\thetaE \over \RtE} = {\pirel \over \AU}.
\end{equation}
Combining (3) and (4), this gives
\begin{equation}
\thetaE  = \sqrt{ {4GM\over c^2}{\pirel\over \AU}}, \qquad\qquad
\RtE  = \sqrt{ {4GM\over c^2}{\AU \over \pirel}},
\end{equation}
where $\pirel = 1/\Dol - 1/\Dos$ is the relative source-lens parallax.
As first pointed out by Gould (2000), these formulae give the physical
parameters ($\pirel, M$) in terms of quantities that are in principle
measurable ($\thetaE, \RtE$).

Fig.~2 shows what happens when the lens (L) is offset from the line
joining observer (O) and source (S). The angular position of the
source from the optic axis is $\thetaS$, while the position of the
image is $\thetaI$. Again, directly from the figure, we see that
$\alpha (\Dos -\Dol) = \Dos(\thetaI - \thetaS)$. Substituting from the
Einstein deflection angle formula~(1), we deduce that
\begin{equation}
\thetaI^2 - \thetaI\thetaS = \thetaE^2.
\end{equation}
This is a quadratic equation for the angular image positions, from
which we deduce that there are two images, henceforth denoted by
I$_{\pm}$. In microlensing, the images are separated by
microarcseconds and so are not resolved. It is useful to introduce
the normalised source position, i.e., $u = {\thetaS}
/\thetaE$. Solving the quadratic, we find that the two images are at
\begin{equation}
{{\btheta}_{\rm I} \over \thetaE} = \pm u_{\pm} {\bf {\hat u}},
\qquad\qquad u_{\pm} = {\sqrt{u^2+4} \pm u \over 2}.
\end{equation}
This is illustrated in Fig.~3, which is a cross-section through the
lens plane. The two images (I$_\pm$), and their centroid (C), lie on
the line joining lens (L) and source (S).  We see that the light
centroid deviates from the true source position. As the lens and
source are in relative motion, the light centroid changes both in
position and brightness as the event progresses.  Microlensing is
therefore detectable both astrometrically and photometrically.

{\it Astrometric microlensing} is the name given to the dance on top
of the parallactic and proper motion caused by a nearby lens (e.g.,
Boden, Shao \& van Buren 1998; Dominik \& Sahu 2000). This is
illustrated in Fig.~4, which shows the right ascension and declination
of a source star in the absence and presence of microlensing.  If an
event is followed astrometrically, then the quantities $\thetaE$ and
$\RtE$ are measurable.  Astrometric microlensing has not been observed
so far, but it will be within the next decade by one of the
astrometric satellites, either the {\it Space Interferometry Mission}
({\it SIM}) or {\it GAIA}, both of which offer microarcsecond
astrometry.  {\it SIM} is a pointing satellite, which will be able to
follow up individual alerted events in a wealth of detail (e.g., Salim
\& Gould 2000). {\it GAIA} is a scanning satellite, which will survey
the whole sky down to $V \approx 20$ and will discover $\sim 25\,000$
astrometric microlensing events (Belokurov \& Evans 2002).

\begin{figure}
\plotone{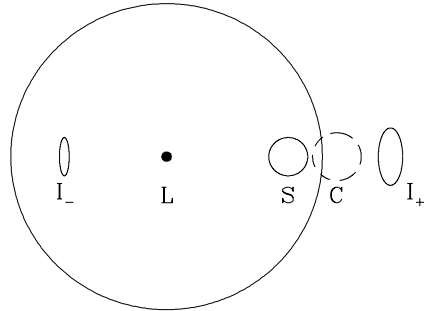}
\caption{This is a slice through the lens plane. The lens (L) lies at
the centre, whilst the true position of the background source is at
S. The two micro-images are at $I_\pm$, and so lie on the line joining
lens and source.  The centroid of the light of the two micro-images is
at C.}
\end{figure}
\begin{figure}
\plotone{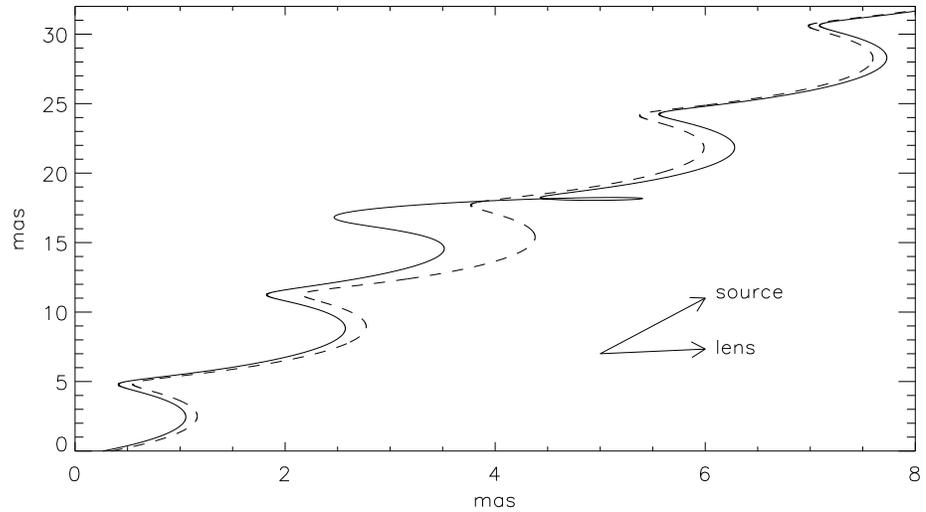}
\caption{The dashed line shows the astrometric path (right ascension
and declination) of an unlensed source, with the yearly parallactic
motion superposed on the proper motion. The full line shows the path
of the light centroid when the source is microlensed by an intervening
dark object [From Belokurov \& Evans 2002].}
\end{figure}

{\it Photometric microlensing} is the brightening and fading of the
source star. In gravitational lensing, surface brightness is
conserved. The magnification of each image $A_{\pm}$ is given by the
ratio of the area of the image to the area of the source, so the total
magnification $A$ can be calculated as
\begin{equation}
A_{\pm} = {\partial \thetaI \over \partial \thetaS} = {u_{\pm}^2 \over
u_+^2 - u_-^2}, \qquad
A = A_+ + A_- = {u^2 +2 \over u \sqrt{u^2 +4}}.
\end{equation}
If the source, lens and observer are all in rectilinear motion, then
$u^2(t) = u_0^2 + (t-t_0)^2/\tE^2$, where $t_0$ is the time of closest
approach, $u_0$ is the (normalised) impact parameter and $\tE$ is the
Einstein crossing time defined as
\begin{equation}
\tE = {\thetaE \over \mu_{\rm rel}},\qquad\qquad \mu_{\rm rel} =
\mu_{\rm l} - \mu_{\rm s}.
\end{equation}
Here, $\mu_{\rm rel}$ is the relative proper motion of the lens.
These equations define the standard Paczy\'nski light curves,
illustrated in Fig.~5. From a photometric microlensing event, we can
only measure $t_0, u_0$ and $\tE$. Of these, only $\tE$ carries any
physical information, being related to the mass, velocities and
distances in a complicated way, namely
\begin{equation}
\tE = {1\over v} \sqrt{4GM \Dol (\Dos-\Dol) \over c^2 \Dos},
\end{equation}
where $v$ is the relative motion at the lens.  Only by statistical
analyses of ensembles of events using Galactic models can physical
information such as the lens masses be extracted.

However, if a microlensing event is monitored photometrically and
astrometrically, then $\thetaE, \RtE, \phi$ (the angle of the
source-lens relative proper motion), $\pi_s$ (the source parallax) and
$\mu_s$ (the source proper motion) are all additionally measurable.
From eq (5), the mass of the lens is then immediately known. The
enormous advantage of astrometric microlensing over photometric makes
its detection a key challenge for the next decade.

\begin{figure}
\plotone{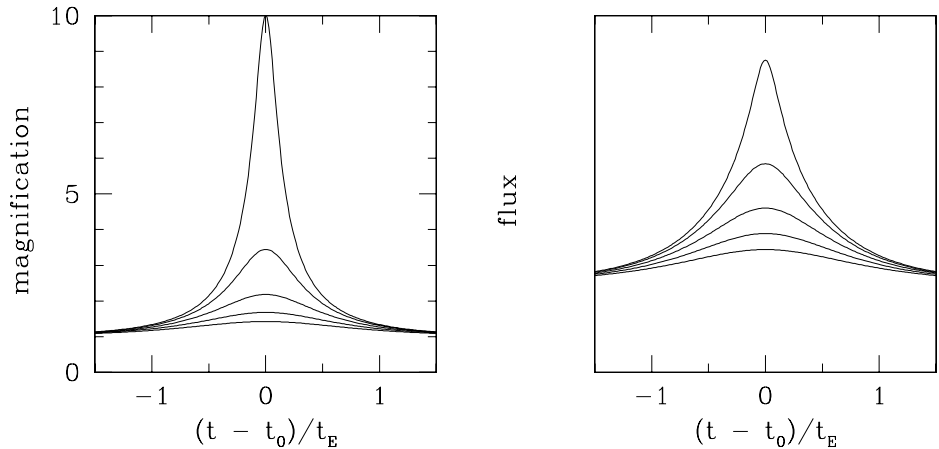}
\caption{Plots of the magnification (left) and the flux (right) for
vanilla microlensing with normalised impact parameter $u_0 = 0.1, 0.3,
0.5, 0.7$ and 0.9. These are often referred to as standard Paczy\'nski
lightcurves.  The height of the curves depends only on the impact
parameter and carries no physical information. The width of the curves
is controlled by the timescale of the event $t_E$.}
\end{figure}

\subsection{Exotic Microlensing}

An exotic microlensing event is one for which we can measure more than
just $t_0, u_0$ and $\tE$ from the lightcurve alone. Though rare, such
events are important as they are the only ones for which additional
information (such as the lens mass or location) can be inferred.

\subsubsection{Parallax Events}

In a parallax event, there are measurable distortions in the
lightcurve caused by the motion of the Earth (e.g., Gould 1992; Hardy
\& Walker 1995).  A parallax event occurs [1] if the Einstein radius
projected onto the observer's plane is roughly the same size as the
Earth's orbit ($\RtE \sim 1$ AU) and [2] if the Einstein crossing time
is reasonably long ($\tE \gta 50$ days), so that the acceleration of
the Earth's motion becomes apparent during the course of the event.
For such an event, the size of the Einstein ring radius projected onto
the observer's plane $\RtE$ is measured by scaling the event against
the size of the Earth's orbit.
 
Fig.~6 shows a beautiful example of a parallax event, namely OGLE
99-BLG-32 (Mao et al. 2002).  This is the longest ever microlensing
event detected to date with $\tE$ = 640 days. It has $\RtE \sim 30$
AU, so that the transverse velocity projected onto the observer's plane
is $\sim 80 \kms$. The degeneracy is only partially lifted by the
parallax effect, so that the lens mass is still not uniquely
determined. However, likelihood fits with Galactic models (e.g., Agol
et al. 2002) suggest that the lens is at least a few solar masses and
probably a black hole.  Bennett et al. (2002) have searched through
the 7 year dataset taken by the MACHO collaboration towards the
Galactic bulge and found six long timescale events with a detectable
parallax signature, of which they reckon five are probable black hole
candidates. Microlensing is the only technique known to us for the
detection of isolated black holes that are not accreting.

Smith et al. (2002) found another very remarkable lightcurve, namely
OGLE 99-BLG-19. This is the first multi-peaked parallax event. The
exceptionally dramatic parallax effect occurs because of the small
relative velocity of the lens projected onto the observer's plane ($\sim
10 \kms$), much smaller than the speed of the Earth around the Sun
($\sim 30 \kms$). Sometimes even the absence of a parallax effect can
be interesting, as it can be used to constrain the location of a
lens. For example, the first event detected towards the Small
Magellanic Cloud (SMC), namely MACHO 97-SMC-1, had a timescale of 123
days (Alcock et al. 1997c) but showed no detectable deviations from a
standard Paczy\'nski curve. This suggests that the lens is not close
to us and tends to favour a location in the SMC itself.

\begin{figure}
\plotone{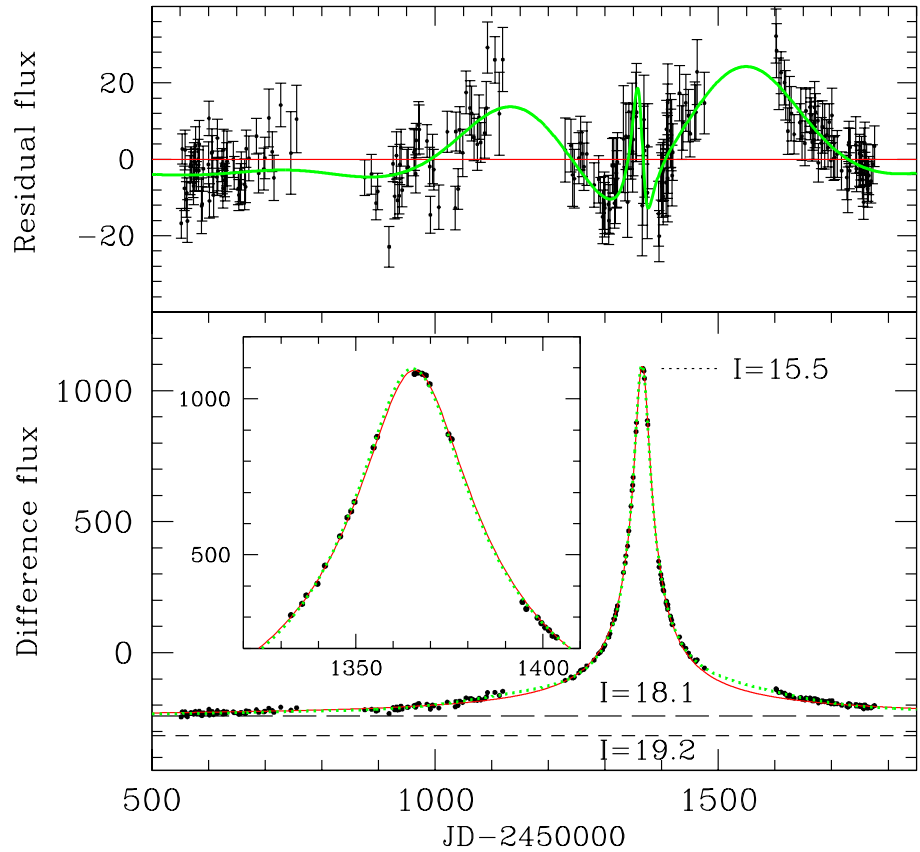}
\caption{This is the lightcurve for OGLE 99-BLG-32, a remarkable event
in many ways. It is the longest known microlensing event to date. The
lower panel shows the difference image flux, together with a standard
and a parallax fit. The upper panel shows the residuals (the observed
flux minus the standard fit). It demonstrates that a standard fit is
unsatisfactory, as it shows systematic discrepancies. However, the
parallax fit nicely reproduces the pattern of the deviations [From Mao
et al. 2002].}
\end{figure}

\subsubsection{Xallarap Events}

Xallarap is a neologism first used in print by Bennett (1998). It is
parallax backwards; the xallarap effect is the converse of the
parallax.  In a xallarap event, there are measurable distortions in
the lightcurve caused by the motion of the binary source (e.g., Han \&
Gould 1997; Dominik 1998).  A xallarap event occurs [1] if the
Einstein radius projected onto the source plane is of the same size as
the binary semimajor axis ($\RhE \sim a$), and [2] if Einstein
crossing time is long compared to the binary period ( $\tE \gta \tau$)
so that the acceleration of the binary becomes apparent during the
course of the event.  For such an event, the size of the Einstein ring
radius projected onto the source plane $\RhE$ is measured by scaling
the event against the size of the binary's orbit (which has to be
inferred by other astrophysical means).
 
An example of a xallarap event is MACHO 96-LMC-2, identified by Alcock
et al. (2001a). There are detectable deviations at the maximum of the
lightcurve which cause a xallarap fit to be preferred to the standard
one. The additional information provided by the xallarap effect
suggests that the lens resides in the Large Magellanic Cloud (LMC)
rather than the Galactic halo. The binary source has a total mass of
$\sim 2$ \msun. In the preferred fit, the primary contributes all the
light, while the secondary is a dark companion.

\subsubsection{Finite Source Size Effects}

When the distance of closest approach is comparable to or smaller than
the stellar radius ($u_0 \lta R_\star$), then the source star can no
longer be regarded as point-like (Witt \& Mao 1994, Gould 1994a). This
typically causes measurable distortions to the peak of the event,
which may be brighter than or fainter than that for a point source.
Another possibility is that the lens may even transit in front of the
disk of the source star, which causes inflection points in the
lightcurve and hence allows a measurement of the crossing time
(Nemiroff \& Wickramasinghe 1994).

An example of microlensing event which shows dramatic deviations from
the standard Paczy\'nski curve due to finite source size effects is
MACHO 95-BLG-30 (Alcock et al. 1997d).  This is caused by the transit
of the lens across the face of the source star and enables measurement
of the lens angular impact parameter in terms of the source size,
namely $\theta_{\rm min} / \theta_\star$ = 0.7. This information,
together with spectroscopic and photometric data, suggests that the
source is an M4 star of radius $\sim 60$ solar radii located on the
far side of the bulge at $\sim 9$ kpc.

\subsubsection{Binary Events}

The most common source of deviation is binarity (e.g., Mao \&
Paczy\'nski 1991).  A point lens lightcurve is defined by just three
parameters, $t_0, u_0$ and $\tE$. A static binary lens is defined by
at least six parameters, namely: (i) the projected separation of the
binary $d$ in terms of the Einstein radius, (ii) the binary mass ratio
$q$, (iii) the Einstein timescale $\tE$ associated with the combined
mass of the binary, (iv) the angle $\alpha$ at which the source
crosses the binary axis, (v) the smallest separation of the source
relative to the center of mass $u_0$ and (vi) $t_0$ which is the time
when $u =u_0$.  Binary microlensing events are important because they
are usually accompanied by caustic crossings. A caustic is a curve in
the source plane which marks the locus of infinite magnification. If a
source passes near or across a caustic, huge changes in magnification
can reveal the angular structure of the source (e.g., Gould 1994a;
Witt 1995). Modelling of the lightcurve then enables the measurement
of $\rho_\star = \theta_\star / \thetaE$.  In other words, the angular
Einstein ring $\thetaE$ is measured by scaling the event against the
angular size of the source $\theta_\star$, which can be determined
from the source flux and colour, together with a colour/surface
brightness relation.  If the observations are sufficiently detailed,
then the intensity variation over the disk of the source star
(limb darkening coefficients) may be inferred.

A spectacular example of a binary caustic crossing event is provided
by MACHO 98-SMC-1.  The source star lies in the Small Magellanic Cloud
(SMC). The event was alerted just after the entry into the caustic,
and the exit from the caustic was intensively monitored by several
groups in five passbands (Afonso et al. 2000). From the lightcurves,
the time taken for the caustic to cross the face of the source can be
deduced. If the distance of the source can be inferred by
astrophysical means, then the projected velocity $\vp$ of the lens at
the source can be computed. The Galaxy halo's optical depth peaks at a
heliocentric distance of $\sim 10$ kpc and the characteristic lens
velocity is $\sim 200 \kms$, giving a projected velocity at the SMC of
$\vp \sim 1000 \kms$. For a lens in the SMC itself, the projection
factor is nearly unity, so $\vp \sim 50 \kms$. For comparison, the
value of the projected velocity measured from the data on MACHO
98-SMC-1 is $\sim 80 \kms$.  So, the additional information provided
by the binary caustic crossing shows that the lens is most likely to
be in the SMC than the Galactic halo itself (Kerins \& Evans
1999). The wealth of detail also enabled limb darkening coefficients
to be computed for five passbands.

Pride of place amongst the exotic events must go to An et al.'s (2002)
analysis of EROS 2000-BLG-5, which was the first time that a microlens
mass was measured. Here, the degeneracy was completely broken by a
combination of exotic effects.  The complex light curve of EROS
2000-BLG-5 has three peaks, two being caused by the entrance and exit
to a caustic and the third by the source's close passage to a
cusp. This tells us that the lens is a binary. The three photometric
peaks allow the source position to be located exactly relative to the
lens geometry at three distinct times, thus enabling the Einstein
radius projected onto the observer's plane to be measured using the
parallax effect ($\RtE \sim 3.6$ AU).  Moreover, finite source size
effects can be measured during the caustic crossings and fix the ratio
of the angular Einstein radius to the angular source size. Given an
estimate of the source size from its position on the colour-magnitude
diagram, this yields the angular Einstein radius ($\thetaE = 1.4$
mas). Referring back to eq. (6), we see that the mass of the lens is
now completely defined. Accordingly, An et al. (2002) measured the
mass as $0.61$ \msun\ and concluded that the lens is a low mass disk
binary (M dwarf) system about 2 kpc from the Sun.  Very recently,
Smith et al. (2003) have also carried a mass determination for a
microlensing event in the OGLE-II database. This lightcurve too shows
both finite source size and parallax effects. The lens mass is $0.05$
\msun, which is interesting as it lies within the brown dwarf
r\'egime. However, in this event, the parallax signature is weak, and
it is possible that the effect is really caused by a binary source. As
Smith et al. (2003) point out, this possibility can be tested by
follow-up spectroscopy.

This completes our description of the fundamentals of vanilla and
exotic microlensing. In the next three sections, we describe
applications of microlensing to studies of dark matter (\S 3), the
structure of the Milky Way and Andromeda galaxies (\S 4) and limb
darkening (\S 5) in turn.


\section{Application I: Dark Matter}

\subsection{Dark Matter Candidates}

The nature of the dark matter haloes surrounding spiral galaxies is a
problem of enormous strategic importance in modern physics and
astronomy.  Baryonic candidates include; (i) black holes, (ii) stellar
remnants, such as neutron stars or white dwarfs, (iii) red dwarfs or
very faint stars, (iv) brown dwarfs, which are stars made from
hydrogen and helium but are too light to ignite nuclear fusion
reactions, (v) Jupiters, which are hydrogenous objects with masses
$\sim 10^{-3}\,\Msun$, (vi) snowballs, which are compact objects with
masses $ < 10^{-3}\,\Msun$ and held together by molecular rather than
gravitational forces and (vii) clouds of molecular hydrogen (e.g, Carr
1994; Evans 2002).  Microlensing searches can detect almost all these
forms of baryonic dark matter, except diffuse clouds of gas.
Non-baryonic candidates include; (i) elementary particles, such as
massive neutrinos, axions or neutralinos, (ii) topological defects in
a gauge field, and (iii) primordial black holes made out of
radiation. Microlensing searches can detect primordial black holes,
but not elementary particles or topological defects.

In the early 1990s, a number of authors (Ashman \& Carr 1988; Thomas
\& Fabian 1990) suggested that cooling flows may have occurred at
cosmological epochs and that galactic haloes may consist of low mass
stars formed in such flows.  There even appeared to be supporting
evidence from star count data in our Galaxy (Richer \& Fahlman 1992).
Red and brown dwarfs were therefore foremost candidates for the dark
matter in galaxy haloes.  In fact, red dwarfs (M and L dwarfs) are the
commonest stars in the Galaxy.  About $90 \%$ of all stars are red
dwarfs.  They have masses between $\sim 0.5$ \msun\ and $\sim 0.08$
\msun, and shine due to hydrogen burning in their cores.  Brown dwarfs
are objects lighter than $\sim 0.08$ \msun. They are too light to
ignite hydrogen. They are brightest when born and then continuously
cool and dim. Near-infrared surveys (DENIS and 2MASS) have been
discovering abundant brown dwarfs since 1997. Reid et al. (1999)
reckoned that the local number density of brown dwarfs is as high as
$0.1$ per cubic pc. In which case, the total number of brown dwarfs
exceeds the total number of all stars in the Galaxy.

In the early 1990s, by contrast, white dwarfs were regarded as rather
improbable dark matter candidates. The main problem is that white
dwarfs have masses in the range $\sim 0.5$ \msun, but are remnants of
stars with masses in the range 1-8 \msun. So, the manufacture of white
dwarfs is necessarily accompanied by the disgorging of substantial
amounts of gas and metals into the ISM, whose presence would surely
have been detectable by now if the dark matter were comprised of
abundant white dwarfs.  It also needs a contrived mass function so as
to avoid leaving large numbers of visible main sequence precursors
still burning today in the halo.

\subsection{The MACHO and EROS Experiments}

The raw data in a microlensing experiment yield a rate (number of
stars microlensed per million stars monitored per year) and a
timescale distribution (number of events with timescales between $\tE$
and $\tE + d\tE$). These observables depend on the distribution of
masses of the lenses and the distribution of proper motions of the
lenses and the sources, all of which are unknown.  At first sight,
therefore, it seems that little definite can be established from
microlensing data. In fact, this is not the case, as a robust quantity
can be calculated from the observables. Suppose there is a threshold
amplification, say $1.34$, above which microlensing can be
detected. This means that the lens lies within an Einstein radius of
the line-of-sight between observer and source for a detectable
event. Let us imagine a tube of circular cross section whose radius is
the Einstein radius (e.g., Griest 1991).  So, the tube attains its
maximum cross-section half-way between observer and source. The
microlensing optical depth $\tau$ is just the number of lenses in this
tube
\begin{equation}
\tau = {\pi\over M} \int_0^{\Dos} \RE^2 \rho(\Dol)\, d\Dol,
\end{equation}
where $M$ is the characteristic mass and $\rho$ is the density of
lenses.  This is independent of the velocities of the lenses and
sources by construction.  It is also independent of the masses of the
lenses (as $\RE \propto \sqrt{M}$). The microlensing optical depth is
a robust quantity depending only on the lens density distribution,
which can be compared against predictions from Galactic models. It has
a natural interpretation as the probability that a given star is being
microlensed. It can also be calculated directly from the data as
a sum over the detected events:
\begin{equation}
\tau = {\pi \over 4} \sum_i {t_{0,i} \over N T \epsilon(t_{0,i})},
\end{equation}
where $N$ is the number of stars monitored, $T$ is the duration of the
experiment, $t_{0,i}$ is the timescale of the $i$th event and
$\epsilon$ is the efficiency as a function of timescale.

Inspired by Paczy\'nski's (1986) suggestion, the MACHO and EROS
experiments began monitoring millions of stellar images in the Large
Magellanic Cloud (LMC) in 1993. From 5.7 years of data, the MACHO
collaboration found between 13 to 17 microlensing events and reckoned
that $\tau \sim 1.2^{+0.4}_{-0.3} \times 10^{-7}$ (Alcock et
al. 2000b). They argued that, interpreted as a dark halo population,
the most likely mass of the microlenses is between 0.15 and 0.9 \msun\ 
and the total mass in these objects out to 50 kpc is found to be
$9^{+4}_{-3} \times 10^{10}$ \msun. This is $\lta 20 \%$ of the halo.
From 8 years of monitoring the Magellanic Clouds, the EROS
collaboration found three microlensing candidates towards the LMC and
one towards the SMC (Lasserre et al. 2000). The EROS experiment
monitors a wider solid angle of less crowded fields in the LMC than
the MACHO experiment, so the two experiments are not directly
comparable. Even though EROS do not analyze their data in terms of
optical depth, it is clear that their results point to a lower value
than that found by MACHO.

\begin{table}[t]
\begin{center}
\begin{tabular}{|l|c|} \hline
\null & \null \\
Optical depth of the thin disk & $0.15 \times 10^{-7}$ \\ 
Optical depth of the thick disk & $0.04 \times 10^{-7}$ \\
Optical depth of the spheroid  & $0.03 \times 10^{-7}$ \\ 
Optical depth of the LMC disk (centre) & $0.53 \times 10^{-7}$ \\ 
\null & \null \\ \hline
TOTAL &$0.75 \times 10^{-7}$ \\ \hline
\end{tabular}
\caption{Inventory of the optical depths of known stellar populations
in the outer Galaxy or the Large Magellanic Cloud (taken from Alcock et
al. 1997b).}
\end{center}
\end{table}

The first question to ask is: have the MACHO and EROS experiments
detected any signal of the dark halo whatsoever?  There are a number
of known stellar populations in the outer Galaxy and the LMC that
contribute to the microlensing optical depth, as listed in Table~1. Of
these, only the optical depth of the LMC remains controversial, with a
number of authors arguing for significantly higher values (e.g., Sahu
1994; Evans \& Kerins 2000). Even using the conservative value given
in Table 1, the microlensing optical depth caused by known stellar
populations is $\sim 0.75 \times 10^{-7}$, which is within $2\sigma$
of the value deduced from the 5.7 yr MACHO observations.  There may
even be hitherto undetected populations in the outer Galaxy or the LMC
-- such as tidal debris (Zhao 1998), the warped outer Milky Way disk
(Evans et al.  1998) or an intervening dwarf galaxy (Zhao \& Evans
2001). Hence, it is possible that the microlensing signal comes
entirely from foreground or background populations and has nothing to
do with the dark halo at all. This viewpoint is supported by the
evidence from the exotic events.  There are now four such events (two
binary caustic crossing events, one long timescale event with no
detectable parallax, one xallarap event) for which the location of the
event can be more-or-less inferred. In all cases, the lens most likely
resides in the Magellanic Clouds.  Most recently of all, there has
been the direct imaging of a lens by Alcock et al. (2001b), revealing
it to be a nearby low-mass star in the disk of the Milky Way.

The second question to ask is: if we do assume that the lenses lie in
the dark halo, what are they? Red dwarfs are ruled out because they
are not seen in sufficient numbers in long exposures of high latitude
wide-field camera {\it Hubble Space Telescope} fields.  Specifically,
less than $1 \%$ of the mass of the halo can be in the form of red
dwarfs (Bahcall et al. 1994; Graff \& Freese 1996).  Brown dwarfs are
ruled out because the timescales of the microlensing events are too
long. By examining different velocity anisotropies and rotation, Gyuk,
Evans \& Gates (1998) showed that the minimum mass of the microlensing
objects must be $\gta 0.1$ \msun, which lies above the
hydrogen-burning limit. So, despite their abundance in the Galaxy,
both brown and red dwarfs cannot be the culprits. White dwarfs remain
possible, at least as regards the timescales. The MACHO collaboration
favoured an explanation in which $\lta 20 \%$ of the dark halo was
built of white dwarfs. However, this has been fiercely contested by
others as being inconsistent with other pieces of astrophysical
evidence.  For example, the existence of multi TeV $\gamma$-rays from
Makarian 501 places a powerful constraint on white dwarfs, as their
progenitors would produce infrared radiation that can interact to
produce electron-positron pairs (Graff et al. 1999). More directly,
overproduction of carbon, nitrogen, deuterium and helium are all in
serious conflict with observations unless the contribution to the
critical density from white dwarfs $\Omega_{\rm WD}$ is less than
$0.003$ (Fields, Freese \& Graff 2000).

\subsection{Conclusions}

Despite such controversies, microlensing has told us a crucial fact
about the Galactic dark halo. {\it Almost all the dark halo is not
built from stellar or sub-stellar compact objects.} A whole swathe of
baryonic dark matter candidates are ruled out as dominant
contributors. The constraints from the spike (or short timescale)
analysis on Jupiters and snowballs are particularly severe. The only
baryonic dark matter contenders that remain possible are supermassive
black holes and clouds of diffuse gas.  Of course, particle dark
matter does not cause microlensing events and remains the most likely
solution of all.
                  

\begin{figure}
\plotone{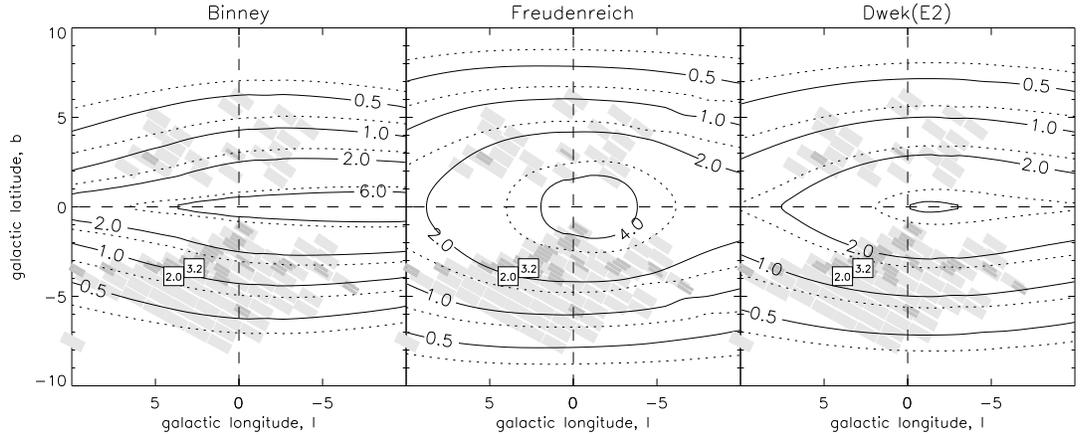}
\caption{Contours of microlensing optical depth to the red clump
giants (in units of $10^{-6}$) in three barred Galaxy models, excluding
(full lines) and including (dotted lines) spirality. The optical
depths reported by Alcock et al. (2000a) and Popowski et al. (2000)
are shown in boxes. Light (dark) gray boxes correspond to EROS (OGLE
II) fields [From Evans \& Belokurov 2002].}
\end{figure}

\section{Application II: Galactic Structure}

\subsection{The Milky Way}

Microlensing surveys towards the Bulge were originally proposed by
Paczy\'nski (1991) and Griest et al. (1991) as a check on the
reliability of the searches towards the LMC. They have now evolved
into uniquely important probes of the mass distribution in the inner
Galaxy. Conventional models of the barred inner Galaxy are derived
from the stellar kinematics, motions of atomic and molecular gas,
starcount data and measurements of integrated light (e.g., Binney et
al. 1991; H\"afner et al. 2000). Such datasets therefore measure
either the light distribution or the gravity field and so they lack
the immediacy of microlensing surveys, which alone measure the mass
distribution directly.

In fact, the early calculations by Paczy\'nski and Griest assumed that
the main lensing population was foreground disk stars. They concluded
that the optical depth to the Galactic Center was of the same order as
that towards the LMC ($\tau \sim 4 \times 10^{-7}$). The first OGLE
and MACHO observational results, although based on small samples,
showed that this was clearly in error (Udalski et al. 1994; Alcock et
al. 1995) and that the optical depth was about an order of magnitude
higher ($\tau \sim 3 \times 10^{-6}$). The important breakthrough was
made by Kiraga \& Paczy\'nski (1994), who first deduced that the main
lensing population was not the disk stars, but the bulge stars
themselves.  Early on, too, it was realised that such high values of
the optical depth doomed purely axisymmetric models of the inner
Galaxy and strongly favoured barred models (Paczy\'nski et al. 1994;
Evans 1994, 1995). Nonetheless, a neat match between the data and the
models continued to elude investigators, until finally Binney,
Bissantz \& Gerhard (2000) raised the alarum with a paper entitled:
``Is Microlensing Compatible with Galactic Structure?''. They argued
that the values of the optical depth for microlensing to bulge sources
-- in particular, the Alcock et al. (2000a) value of $\tau = 3.23
\times 10^{-6}$ -- were so high that they were in conflict with barred
models derived from infrared surveys and gas motions.  Fig.~7 shows
contours of optical depth to bulge sources computed for three popular
models of the inner Galaxy (Evans \& Belokurov 2002). All three models
(Binney, Gerhard \& Spergel 1997; Freudenreich 1998; Dwek et al. 1995)
are derived from the infrared emissivity measured by the DIRBE
instrument on the COBE satellite, but make different corrections for
the distribution of dust. As can been seen, bar models such as Binney
et al.'s and Dwek et al.'s cannot reproduce the high value of $\tau =
3.23 \times 10^{-6}$ at Galactic longitude and latitude ($\ell =
2.68^\circ, b= -3.35^\circ$).  Both these models are highly
concentrated towards the Galactic plane. By contrast, Freudenreich's
model is more massive and swollen, and so gives values of the optical
depth close to the observations. However, in constructing his model,
Freudenreich (1998) masked out most regions close to the Galactic
plane ($|b| < 5^\circ$) as being anomalously reddened by dust. So, the
model is perhaps untrustworthy near the plane, as it depends heavily
on uncertain extrapolations from the outer parts.

\begin{table}[t]
\begin{center}
{\footnotesize
\begin{tabular}{|l|c|c|c|}
\hline
Collaboration  & Location & Optical Depth & Method\\
\null & \null & \null & \null \\
Udalski et al. (1994)  & Baade's Window   & $\sim 3.3 \times 10^{-6}$
& PSF ($f_d=0.0$) \\
Alcock et al. (1995) & ($2.3^\circ,-2.65^\circ$) & $\sim 3.9 \times
10^{-6}$ & PSF ($f_d=0.0$) \\
Alcock et al. (1997a) & ($2.5^\circ,-3.64^\circ$) & $ 3.9^{+1.8}_{-1.2}\times
10^{-6}$ & Red Clump \\
Alcock et al. (2000a) & ($2.68^\circ,-3.35^\circ$) &
$3.23^{+0.52}_{-0.50} \times 10^{-6}$ & DIA ($f_d=0.25$) \\
Popowski et al. (2002) & ($3.9^\circ, -3.8^\circ$) & $2.0 \pm 0.4
\times 10^{-6}$ & Red Clump \\
Popowski (2003) & ($2.2^\circ, -3.2^\circ$) & $2.2^{+0.4}_{-0.4}
\times 10^{-6}$ & DIA ($f_d =0.1$) \\
Sumi et al. (2003)   & ($3.0^\circ,-3.8^\circ$) & $3.40^{+0.94}_{-0.73}
\times 10^{-6}$ & DIA ($f_d=0.25$) \\
Afonso et al. (2003) & ($2.5^\circ, -4.0^\circ$) & $ 0.94 \pm 0.26
\times 10^{-6}$ & Red Clump \\ 
\null & \null & \null & \null \\
\hline
\end{tabular}}
\caption{The microlensing optical depth recorded by various
experimental groups towards locations in the Galactic bulge.  The
method used by each collaboration is also given (DIA = difference
image analysis, PSF = conventional point spread function photometry,
such as SoDoPHOT). If the red clump method is not used, then the
fraction of disk sources $f_d$ must be estimated.}
\end{center}
\end{table}

Table~2 lists the values of the optical depth to sources in the bulge
as measured by a number of investigators.  The best way to measure
this is to use the sub-sample of microlensing events of the red clump
sources only. This is because the red clump stars are known to reside
in the bulge and because they are so bright that the efficiency
depends only on the temporal sampling. Another way is to measure the
optical depth to all sources -- whether by using difference image
analysis (DIA) or conventional point spread function (PSF) photometry
-- and then to correct the total optical depth by the fraction of disk
sources $f_d$. This requires the efficiency of the entire experiment
to be computed, as well as $f_d$ to be estimated from theoretical
models.  Table~2 shows that the optical depth computed from the red
clump stars is lower than than that computed by correcting the total
optical depth (aside from the early values which depend on a handful
of events). The origin of this trend remains unexplained.

The recent, very high result of $\tau \sim 3.4 \times 10^{-6}$ from
the MOA collaboration (Sumi et al. 2003) maintains the inconsistency
between microlensing and galactic structure. This value is in fact
still higher than that reported by Alcock et al. (2000a) and is
measured at a location still further from the Galactic Center. All
three models in Fig.~7 fail to achieve this number by a good margin.
However, there is also a recent, very low result of $\tau \sim 0.94
\times 10^{-6}$ from the EROS collaboration (Afonso et al. 2003) using
the red clump method. In fact, this number can be reproduced rather
easily even by Binney et al.'s model (the least massive and swollen of
the bars in Fig.~7). These two most recent determinations -- amongst
the highest and lowest ever reported -- suggest that the systematics
in these experiments are still not properly understood.
\begin{figure}
\plotone{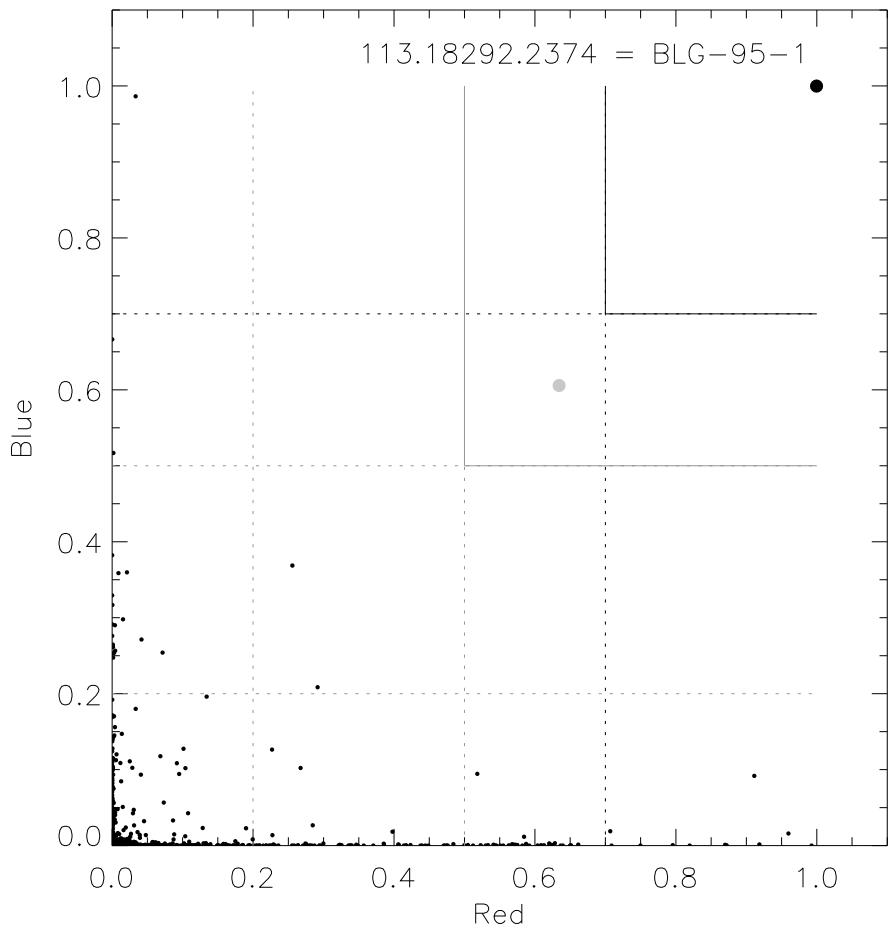}
\caption{The results of processing a tile of MACHO data towards the
Galactic Bulge. The posterior probability of microlensing given the
blue data is plotted against the same probability given the red
data. There is a single microlensing candidate identified by MACHO on
the tile, namely BLG-95-1. It is also cleanly identified by the neural
network as the black spot in the top right hand corner. Note that the
neural network filters almost all the variable stars, which are in the
bottom left hand corner. There is however one pattern marked by a grey
spot which lies in the regime of novelty detection. Close inspection
of the network shows that it is probably a noisy lightcurve of an
eruptive variable [From Belokurov, Evans \& Le Du 2003].}
\end{figure}
\begin{figure}
\plotone{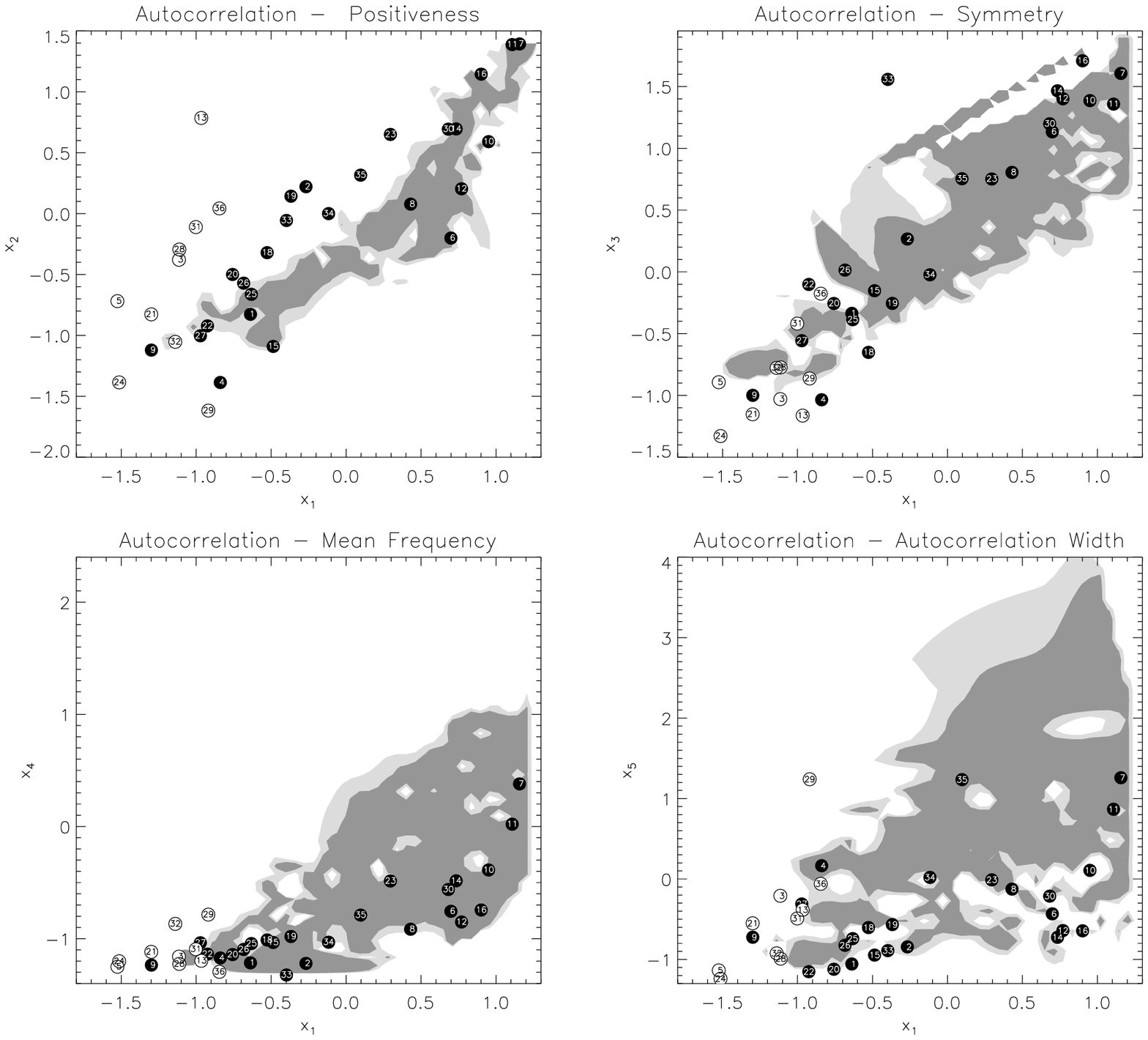}
\caption{The locations of 36 microlensing events (from Alcock et al.
2000a) in the input space ($x_1, x_2, x_3, x_4, x_5$). These
parameters describe the five characteristic features of microlensing
events, namely (i) an excursion from the baseline that is (ii)
positive, (iii) symmetric, (iv) single and controlled by (v) a
timescale. Also shown are the contours of posterior probability of
microlensing for the training sets in the input space.  Light gray
means that the probability is greater than 0.5. Dark gray means that
the probability is greater than 0.9 and corresponds to almost certain
microlensing.  The 36 microlensing events were originally identified
by Alcock et al. (2000a) on the basis of conventional PSF
photometry. The microlensing nature of the events represented by
circles coloured black is corroborated by the network, but those
represented by circles coloured white is not.  There are 11 events not
identified as microlensing when the red data is fed into the neural
network [From Belokurov, Evans \& Le Du 2003].}
\end{figure}

One possibility is that the selection of events is too lax and that
background supernovae and forms of stellar variability are being
inadvertently identified as microlensing. Although the microlensing
phenomenon has a number of characteristic signatures (e.g.,
achromaticity, symmetry, uniqueness), these can be undermined in
heavily crowded fields where blending occurs, or in the case of sparse
and noisy sampling. A recent development has been the exploitation of
neural networks to discriminate between the shapes of microlensing
lightcurves and other contaminants, such as variable stars. Belokurov,
Evans \& Le Du (2003) present a working neural network to identify
microlensing. It has five input neurons, a hidden layer of five
neurons and one output neuron. Microlensing events are characterised
by the presence of (i) an excursion from the baseline that is (ii)
positive, (iii) symmetric, (iv) single and (v) a timescale.  Motivated
by this, five parameters are extracted by spectral analysis from the
lightcurves and fed to the neural networks as inputs. The output of
the network is the posterior probability of microlensing.  For
example, Fig.~8 shows the results of processing all lightcurves in
MACHO tile 18292 of field number 113, which lies towards the Galactic
bulge. This tile contains $\sim 5000$ lightcurves, of which one was
identified by MACHO as a microlensing event. The data are taken at a
site with median seeing of $\approx 2.1^{\prime\prime}$. This means
that the quality of the data is sometimes poor. Each lightcurve is
presented to the neural network, with the red and blue passband data
analysed separately. It would be preferable to analyze the red and
blue data together because most variable stars show colour
differences. However, this option is not viable at the moment because
the publically available colour information on variable stars is still
quite limited.  Fig.~8 shows the results of the deliberations of the
neural network. The probability of microlensing given the blue data is
plotted against the probability given the red data. There is only one
pattern that is unambiguously identified, namely the event designated
by MACHO as BLG-95-1. It is clearly and cleanly separated from the
rest of the patterns as a black circle in the topmost right
corner. There is an additional pattern that has output values $y
\approx 0.6$ for both the red and blue data. This falls within the
regime of novelty detection. It is most probably a form of stellar
variability that the network has not previously met in its training
phase.

Such tests give confidence that the neural network approach is a
fruitful one.  When applied to a larger sets, however, there are some
discrepancies between events identified by MACHO and those identified
by neural networks. For example, Alcock et al. (2000a) identified 36
events towards the Galactic Bulge on the basis of a series of
photometry and colour cuts applied to the lightcurves.  The neural
network finds a total of 19 events identified with a probability $\gta
0.9$ as microlensing in both the red and blue filters. Additionally,
there are 2 events securely identified in the blue data, but not in
the red; there are 6 events identified in the red data, but not in the
blue. Lastly, there are 9 events for which no microlensing signal
whatsoever is detected.  This suggests that the MACHO group's
classification algorithm is itself probably not 100 per cent
efficient.  Fig.~9 shows the contours of probability for the training
set in the input space.  Light gray means that the probability is
greater than 0.5 and corresponds to the formal decision boundary. Dark
gray means that the probability is greater than 0.9 and corresponds to
almost certain microlensing.  The events identified in the red
passband are designated by filled circles, events missed are open
circles. 

In fact, the identification of microlensing events is much more
difficult than usually acknowledged. It probably lies at the heart of
the seeming discord between microlensing and Galactic structure in the
inner Galaxy, as well as the seeming discord between the MACHO and
EROS experiments towards the Large Magellanic Cloud. In order that
microlensing surveys achieve their true status as the most powerful
probe of galactic structure, the identification problem will need to
be much more thoroughly understood than at present.

\begin{figure}
\plotone{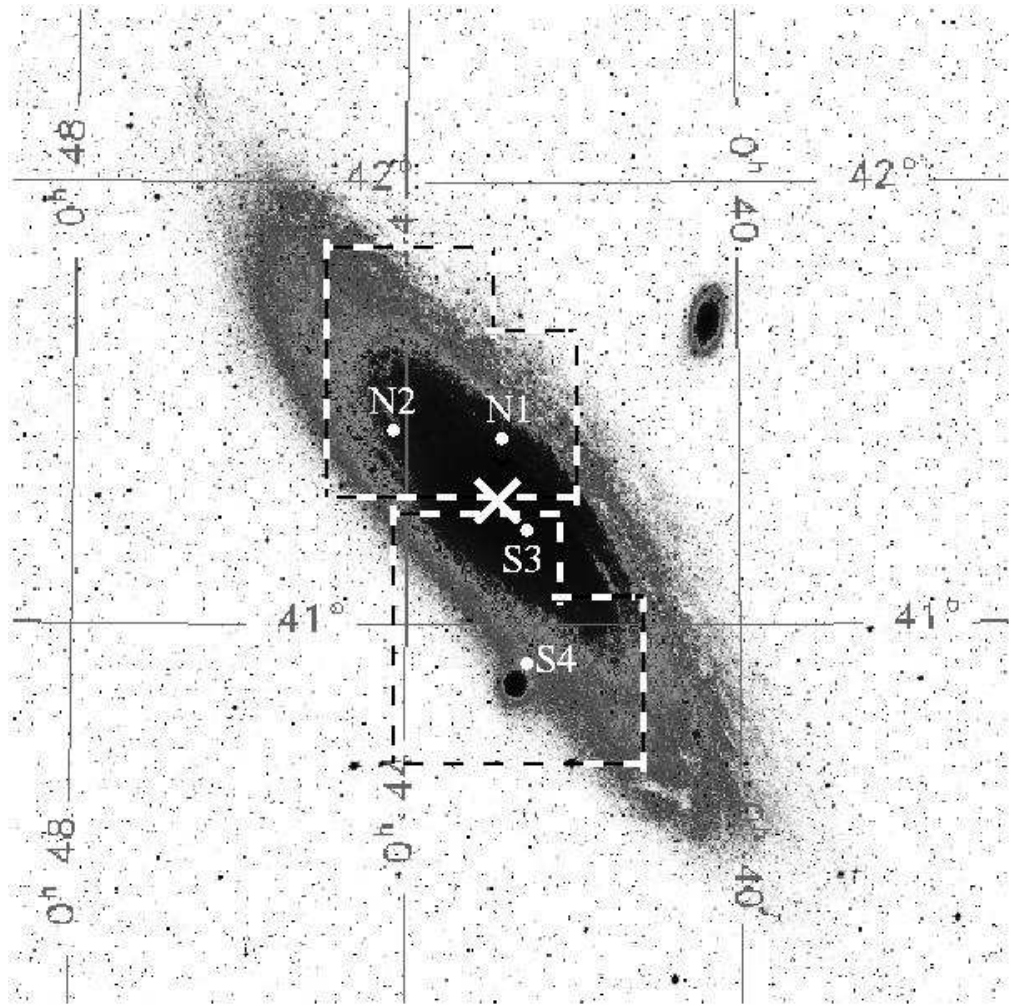}
\caption{The location of the 4 microlensing candidates detected by
POINT-AGAPE towards M31. Also marked are the two INT fields that
straddle the north and south of M31 [From Paulin-Henriksson et
al. 2003].}
\end{figure}
\begin{table}[t]
\begin{center}
{\footnotesize
\begin{tabular}{|l|c|c|c|c|} \hline
reference & $\Delta R$ (mags) & $t_{1/2}$ (days) & $t_E$ (days)
& $A_{\rm max}$ \\ 
\null & \null & \null & \null & \null \\
PA-99-N1  & $20.8\pm 0.1$ & 1.9    & $9.74\pm 0.70$  
& $17.54^{+1.33}_{-1.15}$ \\
PA-99-N2  & $19.0\pm 0.2$ & 25.0   & $91.91^{+4.18}_{-3.83}$ & 
$13.33^{+0.75}_{-0.67}$ \\
PA-00-S3  & $18.8\pm 0.2$ & 2.3    & $12.56^{+4.53}_{-3.23}$ & 
$18.88^{+8.15}_{-5.89}$ \\
PA-00-S4  & $20.7\pm 0.2$ & 2.1    & $128.58 ^{+142.61}_{-72.27}$
& $211^{+16456}_{-120}$ \\ 
\null & \null & \null & \null & \null \\ \hline
\end{tabular}}
\end{center}
\begin{caption}
{Parameters for the 4 POINT-AGAPE candidates. Here, $\Delta R$ is the
magnitude (Johnson/Cousins) of the maximum source flux variation,
$t_E$ is the Einstein timescale, $t_{1/2}$ is the full-width
half-maximum and $A_{\rm max}$ is the maximum amplification. All these
events have very high amplification and short full-width half-maximum
timescale [From Paulin-Henriksson et al. 2003].}
\end{caption}
\end{table}

\subsection{The Andromeda Galaxy}

The Andromeda Galaxy (M31) is now the subject of intense scrutiny by a
number of groups (e.g., Auri\`ere et al. 2001; Riffeser et al. 2001;
Calchi-Novati et al. 2001; Crotts et al. 2001; Paulin-Henriksson et
al. 2002, 2003).  Conventional microlensing is limited to the only
three galaxies in which there are large numbers of resolved stars (the
Milky Way, the LMC and the SMC). In M31, the potentially lensed stars
are much fainter than the integrated light from all the stars within
the seeing disk. So, the observed quantity is the lightcurve
associated with the flux on a pixel or super-pixel. This technique is
known as pixel lensing (e.g., Gould 1996a).  It is no longer possible
to measure the unlensed fluxes of the individual sources, nor the
timescales of individual events.  Rather, an event is characterised by
its full-width half-maximum timescale $t_{1/2}$ which is only crudely
related to the mass of the lens.  Pixel lensing represents the
ultimate limit of scientific detection because not only is the lens
invisible, but so in essence is the source (because it cannot be
distinguished from other stars in the same pixel). The success of the
technique rests upon the variation in the brightness of the source
manifesting itself as a variation in the local surface brightness.
This is potentially a very powerful technique, which offers ``the key
to the Universe'' (Gould 1996b) or at least the key to the nearest 50
Mpc (Binney 2000).

The POINT-AGAPE collaboration (e.g., Auri\`ere et al. 2001; Kerins et
al. 2001) has been conducting a major program of observations using
the Isaac Newton Telescope Wide-Field Camera from 1999 to 2002. They
monitored nightly two fields in two colours near the central bulge of
M31, as shown in Fig.~10. The two fields are each $34^\prime \times
34^\prime$ and provide a grand total of 256 million pixel
lightcurves. The original impetus for the POINT-AGAPE experiment was
to detect an asymmetry in the gradients of microlensing events between
the near or northern side and the south or furthest side (Crotts 1992,
Baillon et al. 1993). This effect arises because the disk of M31 is
highly inclined ($i = 77^\circ$). An asymmetry is produced if most of
the microlensing events are caused by compact objects in a spherical
halo, as lines of sight to the further, southern side are longer than
lines of sight to the nearer, northern side. This asymmetry is not
caused by stellar lenses in M31 or foreground lenses in the Milky Way.

As the POINT-AGAPE data are taken on different nights with different
photometric conditions, the challenge in the experiment is to
eliminate all sources of variation extrinsic to the source. The pixel
method (Ansari et al. 1997, 1999; Paulin-Henriksson 2002) has been
developed to cope with the measurement of flux changes of unresolved
stars in the face of seeing variations. After geometric alignment,
the current frame is photometrically corrected to the reference image
\begin{equation}
{\overline \phi}_{\rm ref} = a {\overline \phi}_{\rm
cur} + b.
\end{equation}
Here, $a$ is the ratio of absorptions (due to variations in airmass or
atmospheric transmissions) and $b$ the difference in sky backgrounds.
The ${\overline \phi}$ values refer to the median flux on a pixel
computed using a running window of size $41 \times 41$ pixel. The
parameters $a$ and $b$ are calculated using the means and dispersion
of the current frame as compared to the reference image
\begin{equation}
\sigma^2_{\rm ref} = a^2 \sigma^2_{\rm cur}, \qquad\qquad
\langle \phi_{\rm ref} \rangle = a \langle \phi_{\rm cur} \rangle + b.
\end{equation}
Here, the means and dispersions are taken over windows of $500 \times
500$ pixels, so as to render any photon noise negligible.

In the pixel method, which is a simple but effective form of
difference imaging, each elementary pixel is replaced by a super-pixel
centered upon it. Each super-pixel is a square of 7 $\times$ 7
pixels. The size of the super-pixel must be chosen empirically so as
to be large enough to cover the whole seeing disk, but not so large
that any variation is diluted.  If each pixel were weighted with the
point spread function, whose width was allowed to vary with the
seeing, then this method would be equivalent to a difference image
analysis, as utilised by Tomaney \& Crotts (1996).  The simpler
super-pixel method corrects for the different loss of flux in the
changing wings of the PSF with the changing seeing by using an
empirical ``seeing stabilisation'' (e.g, Paulin-Henriksson 2002). This
is deduced by looking at the correlation between the differences in
the super-pixel flux and its median on the current and reference
image, namely
\begin{equation}
\phi_{\rm cur} - {\overline \phi}_{\rm cur} = 
(1 + \alpha)\left( \phi_{\rm ref} - {\overline \phi}_{\rm ref} \right)
+ \beta.
\end{equation}
Here, ($\alpha,\beta$) are estimated by using all the super-pixels on
the current frame. They are then used to correct the current super-pixel
flux to the reference seeing, viz
\begin{equation}
\phi = {a \left(\phi_{\rm cur} - {\overline \phi}_{\rm cur} \right) -
\beta \over \alpha +1} + {\overline \phi}_{\rm ref}.
\end{equation}
This gives the photometrically and geometrically aligned, stable flux.
Although crude, this procedure strikes a nice balance between
computational efficiency and optimal signal-to-noise, with the
resulting noise level approaching the photon noise limit.

The POINT-AGAPE collaboration has thus far processed the first two
years of data and has a list of 362 candidate microlensing events
(Paulin-Henriksson et al. 2003). These pass the sequence of cuts
required to isolate candidate events. However, many of these are
probably variable stars and will require additional baseline data
before they can be distinguished from microlensing.  For the moment,
POINT-AGAPE have restricted themselves to high amplitude events with
full-width half-maximum timescales $t_{1/2}$ shorter than 25
days. These cuts eliminate almost all of the troublesome long-period
Mira variables that are the most serious contaminants in this
experiment.  This leaves four, robust high signal-to-noise events
(PA-99-N1, PA-99-N2, PA-00-S3, PA-00-S4), whose characteristics are
listed in Table~3 and whose locations are marked on Fig.~10. Here, 99
and 00 designate the year in which the event reached maximum, while N
and S indicate whether the event occurred in the northern or the
southern field.

Remarkably, all the events that have thus far been discovered can
reasonably enough be ascribed to stellar lenses. The projected
positions of PA-99-N1 and PA-00-S3 lie within the bulge of M31, where
lensing by stars in M31 overwhelmingly dominates over lensing by
objects in the halo.  The projected position of PA-00-S4 lies very
close to the centre of the foreground elliptical galaxy M32. The
detailed analysis of this event by Paulin-Henrikkson et al. (2002)
suggests that the source star lies in the M31 disk, but that the lens
most probably resides in M32 itself. For example, the optical depth to
lensing by M32 stars is $\tau \sim 1.4 \times 10^{-6}$, which is
roughly twice as big as the optical depth to compact, dark objects in
M31's halo (assuming Alcock et al.'s (2000b) value of 20 \% as the
fraction in such objects).  The event PA-99-N2 lies in the disk $\sim
22^\prime$ from the centre of M31. However, its Einstein crossing time
is $\sim 92$ days, making it the longest event so far discovered in
the direction of Andromeda.  A microlensing event with a short
Einstein crossing time far out in the disk would be an unambiguous
candidate for a dark halo lens. Given the long timescale, the most
likely interpretation is that the lens is also a disk star, and that
this is an example of disk-disk lensing (e.g., Gould 1994b). The
optical depth to disk-disk lensing at this location is $\sim 10^{-7}$,
which is of the same order as the M31 halo under the Alcock et al. 20
\% hypothesis.

What is intriguing is that none of the existing candidates seemingly
implicates a lens in M31's halo. Rather, they seem to suggest that the
main lensing populations coincide with the known stellar populations.
If this trend persists, then the experiments towards M31 have the
potential to test whether mass traces light in the M31 bulge and
disk. Such hypotheses are frequently used in galactic modelling, but
have so far never been checked.

\section{Application III: Limb Darkening}

Limb darkening is the name given to the darkening at the rim of the
stellar disk. It is familiar from optical images of the Sun, in which
context it has been extensively studied.  It happens because photons
on lines of sight towards the rim emanate from less deep, and hence
cooler, layers of the Sun than photons on lines of sight towards the
centre. Measurements of limb darkening in stars other than the Sun
would provide a useful check on theories of stellar
atmospheres. However, such measurements are hard to carry out with
conventional techniques.

In a binary lens, a caustic is an extended structure. If the source
passes near or across the caustic, drastic changes in the
magnification can reveal the finite size and the surface brightness
profile of the source.  This opens up the possibility of studying limb
darkening with gravitational microlensing, as envisaged originally by
Bogdanov \& Cherepaschuk (1995), Witt (1995) and Valls-Gabaud (1998).
This technique has been spectacularly exploited in recent years by the
PLANET collaboration.

The phenomenon of limb darkening is normally parametrised according to
either a linear law
\begin{equation}
S_\lambda(\vartheta) = {\overline S_\lambda} \left[
(1- \Gamma_\lambda) + {3 \Gamma_\lambda \over 2} \cos \vartheta
\right],
\end{equation}
or a square-root law
\begin{equation}
S_\lambda(\vartheta) = {\overline S_\lambda} \left[ (1- \Gamma_\lambda
- \Lambda_\lambda) + {3 \Gamma_\lambda \over 2} \cos \vartheta
+{5 \Lambda_\lambda \over 4} \cos^{1/2} \vartheta
\right].
\end{equation}
In these formulae, $S_\lambda$ is the surface brightness of the star
as a function of $\vartheta$, which is the angle between the normal to
the stellar surface and the line of sight. Additionally,
$\Gamma_\lambda$ and $\Lambda_\lambda$ are the limb darkening
coefficients, while ${\overline S_\lambda}$ is the mean surface
brightness. This is related to the total flux received $F_\lambda$ via
${\overline S_\lambda} = F_\lambda/(\pi \theta_\star^2)$ where
$\theta_\star$ is the angular radius of the star.  Depending on the
quality of the data, it may be feasible to extract either one
($\Gamma_\lambda$) or two ($\Gamma_\lambda$ and $\Lambda_\lambda$)
limb darkening coefficients.  Of course, the coefficients are a
function of the waveband of observation.

A source inside a caustic will be imaged into 5 images; outside the
caustic it will be imaged into 3 images. At the caustic, 2 images
appear or disappear. These images are infinitely magnified. In the
immediate neighbourhood of a caustic, the magnification of the two new
images diverges as (e.g., Schneider \& Weiss 1986)
\begin{equation}
A \propto \left( {1 \over \Delta u_\perp} \right)^{1\over 2} H( \Delta u_\perp ),
\end{equation}
where $\Delta u_\perp$ is the perpendicular separation of the source
from the caustic in units of the angular Einstein radius $\thetaE$,
and $H$ denotes the Heaviside step function. Thus, the magnification of an
extended, limb darkened source is just
\begin{equation}
A_\lambda = {1\over F_\lambda} \int \int_{\cal D} d^2 \vartheta A
(\vartheta) S_\lambda(\vartheta),
\end{equation}
where $F_\lambda$ is the total flux. By substituting in the limb
darkening laws [eqs. (17) or (18)], the angular integration can be
performed analytically to yield (e.g., Appendix B of Albrow et
al. 1999b)
\begin{equation}
A = \left( {1\over \rho_\star^{1/2}} \right) \left[ G_0(- \Delta u_\perp
/\rho_\star ) + \Gamma_\lambda F_{1/2} (- \Delta u_\perp /\rho_\star )
+ \Lambda_\lambda F_{1/4} (- \Delta u_\perp /\rho_\star )\right],
\end{equation}
where $\rho_\star = \theta_\star/ \thetaE$ and $G_0, F_{1/2}$ and
$F_{1/4}$ are known functions, specifically
\begin{eqnarray}
G_n(\eta) &=& {1 \over B (n+3/2,1/2)} \int_{\max(\eta,-1)}^1
{dx (1-x^2)^{n+1/2}\over (x-\eta)^{1/2}} H(1-\eta), \nonumber \\
F_n(\eta) &=& G_n(\eta) - G_0(\eta).
\end{eqnarray}
Here, $B(x,y) = \Gamma(x)\Gamma(y)/\Gamma(x+y)$ is Euler's Beta
function.  Therefore, by decomposing into basis functions the
magnification changes of the source as it crosses the caustic , the
limb darkening coefficients $\Gamma_\lambda$ and $\Lambda_\lambda$ can
be extracted.

There are six parameters of a static binary lens. In addition, there
are also the source flux, the background flux and the limb darkening
coefficients for each waveband. This leads to a $\chi^2$ minimization
in at least a nine-dimensional parameter space. In practice, it is
easier to proceed by transforming from position-magnification space to
time-flux space assuming rectilinear motion of the source relative to
the lens. This leads to an analytic approximation to the shape of the
caustic crossing. The caustic crossing fit then constrains the search
for a full solution to a four-dimensional submanifold of the whole
nine-dimensional space (Albrow et al. 1999b).

Microlensing is the only technique available to us for studying the
limb darkening of distant stars. The recent years have seen limb
darkening coefficients measured for two K giants in the bulge (Albrow
et al. 1999a, Albrow et al. 2000), for a late G or early K sub-giant
in the bulge (Albrow et al. 2001) and for an A dwarf in the SMC
(Afonso et al. 2000).  Theories of stellar atmospheres predict limb
darkening laws for different types of stars. Typically the results
seem to show good agreement with theoretical predictions in the V
band, but poorer agreement in the I band -- as, for example, in OGLE
99-BLG-23 studied by Albrow et al. (2001).

\section{The Future}

In this concluding final section, we suggest four projects for the
next decade.

\subsection{K band Microlensing Towards the Bulge}

Microlensing surveys in the K band towards the Bulge would be
extremely valuable (Gould 1995; Evans \& Belokurov 2002). This is all
the more true given the capabilities of the new generation of survey
telescopes.  For example, VISTA~\footnote{http://www.vista.ac.uk} has
a field of view of 0.25 square degrees in the K band.  Assuming that
the seeing is $0.8''$ in Chile and scaling the results of Gould
(1995), then we estimate that VISTA will monitor $\sim 1.5 \times
10^6$ stars in a single field of view for crowding-limited K band
images towards the Bulge. This means that we are probing the
luminosity function down to $K \sim 16$, assuming 3 magnitudes of
extinction.  We estimate that photometry accurate to $3 \%$ for a $K
\sim 16$ star will take about 1 minute on VISTA. Hence, a K band
survey of a $5^\circ \times 5^\circ$ field close to the Galactic
Center will take about 1.5 hours of time every night. This makes a K
band microlensing survey of the inner Galaxy an attractive and
feasible proposition with VISTA.

The scientific returns of a K band microlensing survey towards the
Bulge will be substantial. First, it will provide new and reliable
estimates of the microlensing optical depth for many locations
throughout the Bulge, rather than the isolated windows available in
the optical bands.  Second, the shapes of the contours of optical
depth -- the microlensing maps -- will enable us to discriminate
between bar models, such as those highly concentrated towards the
Galactic plane (like Binney et al.'s) or those that are diffuse and
swollen (like Freudenreich's).

\subsection{Pixel Lensing Towards M33}

M33 is a low luminosity spiral galaxy in the Local Group.  From the
point of view of dark matter studies, it is an interesting target, as
it is known that the dark matter content of low luminosity and dwarf
galaxies is different from that of big bright galaxies (e.g., Evans
2000). From the behaviour of the rotation curve near the centre, it is
clear that dark matter must dominate even the central parts of M33
(Toomre 1981). This is very different from both the Milky Way and M31,
which are dominated by luminous matter within the inner few kpc.

The VLT Survey Telescope (VST) has a pixel size of 0.24 arcsec/pixel
and a field of view of 1 square degree. VST is likely to see first
light in 2003. The likely pixel lensing rate can be crudely estimated
as (see e.g., equation (19) of Ansari et al. 1997)
$$N_{\rm ev} \sim 160 \times
10^{-0.2(\mu_{\rm gal} - \mu_{\rm M31})}
\biggl({ \Omega_{\rm gal} \over 1 {\rm deg}^2}\biggr) 
\biggl({{\rm season}\over 6\ {\rm months}} \biggr)
\biggl( { {\rm seeing}\over 1.5''}\biggr)^{-1/2},$$
where the normalisation has been determined by Kerins et al.'s (2000)
simulations for the campaign on the INT WFC towards M31. Here,
$\Omega_{\rm gal}$ and $\mu_{\rm gal}$ are the target galaxy's solid
angle and mean surface brightness respectively.  For M33, $\mu_{\rm
gal}$ is 23.8 mag arcsec${}^{-2}$ and $\Omega_{\rm gal}$ is 0.6 ${\rm
deg}^2$.  Paranal in Chile, where the VST is based, has an average
seeing of $0.6^{\prime\prime}$.  We take 21.5 mag arcsec${}^{-2}$ as
the average surface brightness of M31 within the INT fields. Using our
formula, this means that for M33, there will be about 50 events per
season for a halo full of substellar compact objects. Taking Alcock et
al.'s (2000b) baryon fraction of $20 \%$ as applicable, then there
will be $\sim 10$ events per season.

Low luminosity spiral galaxies have very different properties to
bright spiral galaxies like the Milky Way. The MACHO and EROS
experiments have demonstrated that the substellar compact objects are
not the dominant contributor to the Milky Way's dark halo. However,
this conclusion cannot be extended to low luminosity galaxies like M33
without further experiments. Hence, a pixel lensing survey of M33 is a
very worthwhile project in the context of dark matter science.

\subsection{Polarimetry of Microlensing Alerts}

A number of theoretical studies have examined polarization changes
during microlensing (Simmons, Willis \& Newsam 1995; Agol 1996;
Belokurov \& Sazhin 1997), but no attempts have been made to detect
this phenomenon observationally yet. In microlensing, the total flux
is partially polarized and the plane of polarization is perpendicular
to the plane joining the centres of the source and the lens. As the
lens and source are in relative motion, the plane of polarization
rotates during the course of the event.  For a point lens, the
polarization has a magnitude of at most $0.1 \%$.  However,
polarizations as high as $1 \%$ can be achieved if the star crosses a
caustic in a binary lens.  By comparison, the first detection of limb
polarization in Algol was at a magnitude of $0.004 \%$ (Kemp et
al. 1983). Hence, the effect is well within the grasp of current
instruments.

The measurement of variable polarization yields the Einstein radius of
the lens (if the radius of the star is known or can be estimated) and
the velocity direction of the lens projected onto the sky.  For a
binary lens event, studies of the polarization give the position angle
of the binary as well.  Stars with high surface temperature are the
most promising candidates for observing polarization as electron
scattering dominates the opacity (Chandrasekhar 1960; Agol 1996).
Measurements of polarization can provide confirmation of the
microlensing nature of an event, and enable theoretical calculations
of polarization in model stellar atmospheres to be checked and
calibrated. Most importantly, the additional information provided by
polarimetry for some exotic events may lead to determinations of the
mass of the lens. It would therefore be worthwhile to follow up a
subsample of, say, $\sim 50$ microlensing alerts with polarimetry.

\subsection{Astrometric Microlensing with {\it GAIA}}

{\it GAIA}~\footnote{http://astro.estec.esa.nl/gaia} is the European
Space Agency satellite now selected as a Cornerstone 6 mission.  It is
the successor to the pioneering {\it Hipparcos} satellite, which flew
from 1989 to 1993.  {\it GAIA} is a survey satellite that provides
multi-colour, multi-epoch photometry, astrometry and spectroscopy on
all objects brighter than $V\approx 20$ (e.g., ESA 2000; Perryman et
al. 2001). The dataset is gigantic, as there are over a billion
objects in our Galaxy alone brighter than 20th magnitude. A small
fraction of the objects monitored by {\it GAIA} will show evidence of
microlensing. {\it GAIA} can observe photometric microlensing by
measuring the amplification of a source. However, {\it GAIA} is
inefficient at discovering photometric microlensing events, as the
sampling of individual objects is relatively sparse (there are a
cluster of observations once every two months on average).

{\it GAIA} is better at detecting astrometric microlensing.  The
all-sky source-averaged astrometric microlensing optical depth is
$\sim 10^{-5}$, which is over an order of magnitude greater than the
photometric microlensing optical depth. There are two main
difficulties facing {\it GAIA} in exploiting this comparatively high
probability. First, the astrometric accuracy of a single measurement
by {\it GAIA} depends on the source magnitude and degrades at
magnitudes fainter than $V \approx 15$.  Second, {\it GAIA} provides a
time-series of one-dimensional astrometry by scanning great circles on
the sky. The observed quantity is the CCD transit time for the
coordinate along the scan. This is the same way the {\it Hipparcos}
satellite worked (ESA 1997). From the sequence of these
one-dimensional measurements, the astrometric path of the source,
together with any additional deflection caused by microlensing, must
be recovered.

Simulations by Belokurov \& Evans (2002) suggest that $\sim 25000$
sources will exhibit astrometric microlensing events during the course
of the 5 year mission.  The cross-section for astrometric microlensing
favours nearby lenses.  The most valuable events are those for which
the Einstein crossing time $\tE$, the angular Einstein radius
$\thetaE$ and the relative parallax of the source with respect to the
lens $\pi_{\rm sl}$ can all be inferred from {\it GAIA}'s
datastream. The mass of the lens then follows directly. If the source
distance is known -- for example, if {\it GAIA} itself measures the
source parallax -- then a complete solution of the microlensing
parameters is available. Of these quantities, it is the relative
parallax that is the hardest to obtain accurately. Belokurov \& Evans
(2002) used a covariance analysis to follow the propagation of errors
and establish the conditions for recovery of the relative
parallax. This happens if the angular Einstein radius $\thetaE$ is
large and the Einstein radius projected onto the observer's plane
$\RtE \sim 1$ AU so that the distortion is substantial. It is also
aided if the source is bright so that {\it GAIA}'s astrometric
accuracy is high and if the duration of the astrometric event is long
so that {\it GAIA} has time to sample it fully. These conditions
favour still further lensing populations that are close (within $\sim
1$ kpc). Monte Carlo simulations suggest that {\it GAIA} can recover
the mass of the lens to good accuracy for $\sim 10 \%$ of all the
events. Astrometric microlensing can detect objects irrespective of
their luminosity and so is sensitive to completely dark populations
like isolated neutron stars and black holes.  This provides an
excellent way of taking a census of the masses of objects in the local
solar neighbourhood.  Astrometric microlensing with {\it GAIA} will be
the best way to measure the local mass function.

\acknowledgements 
NWE thanks the Royal Society for financial support and all his
colleagues for numerous stimulating discussions, and especially Shude
Mao for permission to reproduce Figure 6.

\label{page:last}
\end{document}